\documentclass[journal,a4paper]{IEEEtran}

\usepackage[dvips]{graphicx}	%Permette di lavorare con le immagini
\usepackage{mathtools}			%Permette di unire simboli e creare ad esempio il valore assoluto
\usepackage{bm}					%Per il corsivo grassetto nelle formule matematiche (\br{})
\usepackage{flushend}%Per bilanciare la lunghezza delle colonne dell'ultima pagina
\usepackage{paralist}			%Permette di inserire un elenco numerato sulla stessa linea
\usepackage{multirow}			%Per scrivere su più righe in una tabella
\title{A Synthetic MIMO PLC Channel Model}

\author{
\IEEEauthorblockN{Alberto Pittolo$^1$ and Andrea M. Tonello$^2$\\}
\IEEEauthorblockA{$^1$University of Udine, Udine, Italy -- e-mail: alberto.pittolo@uniud.it\\
				  $^2$Alpen-Adria-Universit{\"a}t, Klagenfurt, Austria -- e-mail: andrea.tonello@aau.at}
\thanks{This work has been presented at the 2016 IEEE 20th International Symposium on Power Line Communications and its Applications (ISPLC) -- Bottrop, Germany, 20-23 March 2016, in the recent results session.}
}

\begin{document}

\maketitle

%\vspace{-10pt}
\begin{abstract}
The huge and increasing demand of data connectivity motivates the development of new and effective power line communication (PLC) channel models, which are able to faithfully describe a real communication scenario. This is of fundamental importance since a good model represents a quick evaluation tool for new standards or devices, allowing a considerable saving in time and costs. The aim of this paper is to discuss a novel top-down MIMO PLC synthetic channel model, able to numerically emulate a real PLC environment. First, the most common channel modeling strategies are briefly described, highlighting strengths and weaknesses. Afterwards, the basic model approach is described considering the SISO scenario. The implementation strategy is then extended to the MIMO case. The validity of the proposed model is proved making a comparison between the simulated channels and channels obtained with measurements in terms of both performance and statistical metrics. The focus is on the broadband frequency spectrum.
\end{abstract}


%\IEEEpeerreviewmaketitle

\section{Introduction}
In recent years, we have assisted at the huge demand of data connectivity by users, but also by the widespread use of ``smart'' devices that need to exchange information. Within this context, the exploitation of the wireless technology alone will not be enough. Another attractive solution is represented by the transmission of information over the power delivery infrastructure, named power line communication (PLC), which plays an important role and is becoming more and more popular. In particular, PLC is very useful for applications such as: remote metering, local area networks, home networking and automation, as well as in the Smart Grid context. A first industrial standard for indoor PLC applications was developed by the HomePlug (HP) Alliance for devices operating in the 2--28 MHz band. The latest version, namely HPAV2 \cite{HPAV2}, promises to achieve up to 2 Gbps extending the bandwidth up to 86 MHz, exploiting multiple-input multiple-output (MIMO) transmission, using precoding and adaptation, together with a number of advanced techniques at the MAC layer. Moreover, the first worldwide broadband standard was defined by IEEE (standard P1901) \cite{IEEE_P1901}, followed by ITU (G.9960 known as G.hn) \cite{ITU_Ghn}.

In order to provide effective channel models, suitable for the design of innovative devices and for the definition of new standards, it is of fundamental importance to assess, and then be able to reproduce, all the channel properties and the specific features of the different environments. This is even more true when considering a PLC network, where different wiring structures and network topologies, as well as line discontinuities and unmatched loads, imply an extreme variability of the channel. Thus, a thorough knowledge of the transmission medium is needed in order to develop reliable channel models, able to faithfully describe a real network.

This paper considers both the in-home SISO and MIMO scenarios in the broadband frequency range. First, the most common channel modeling strategies are defined and compared, assessing differences and similarities. Then, the attention is moved towards the top-down modeling strategy. The aim is to exploit all the available information concerning the channel properties in order to develop a simple and effective top-down synthetic channel model, suitable for any SISO or MIMO PLC scenario. The modeling strategy is initially described for the SISO case. Afterwords, it is extended to the MIMO context. Both the procedures are validated through numerical results. The conclusions follow.

\section{Channel Modeling}
The characterization of the PLC channel, its properties and the relationship between its metrics, is of fundamental importance in order to develop a fair and effective channel model. The reasons for the development of new and simple models is mainly related to the researchers, developers and manufacturers needs, that demand a quick and cost effective way to test the new devices. In this perspective, two different channel modeling strategies can be approached, namely bottom-up and top-down. Each strategy can follow a deterministic or a statistical approach and has its own strengths and weaknesses.

The bottom-up approach relies on the transmission line theory in order to describe the channel transfer function, representing the complex network structure via ABCD or scattering parameter matrices. A statistical extension to this approach was firstly proposed in \cite{BottomUp} and then extended in \cite{BU_Ton}, which offered a realistic statistical description of in-home networks. Unfortunately, this approach requires a complete network topology knowledge in terms of wiring cables and loads, leading to computationally intensive procedures.

Conversely, following a top-down strategy, the channel response is obtained by fitting a given parametric analytic function with the data coming from the experimental measures. In particular, in \cite{MPM} the CFR was modeled taking into account the multipath nature of the signal propagation, as well as the cables losses. Based on the model provided in \cite{MPM}, a statistical extension to the top-down approach was firstly presented in \cite{TD_Ton}. Although the model discussed in \cite{MPM} belongs to the top-down family, it still embeds some physical concepts and can be further simplified. Indeed, according to recent results discussed in \cite{IEICE}, the CFR can be easily and directly modeled via a synthetic channel model that generates amplitudes and phases as correlated complex random variables. Hence, the top-down approach represents a promising technique to develop simple and effective channel models, as discussed in the following.

\subsection{A SISO Synthetic Channel Model Approach}
\label{SISOmodel}
In this section, a brief description of the synthetic channel model presented in \cite{IEICE} is discussed for the in-home SISO case. The focus is on the broadband frequency range, e.g., in the 1.8--100 MHz band. The main idea is to consider the CFR in amplitude and phase as $H(f)=A(f)e^{j\varphi(f)}$ and to exploit their joint statistics.

\begin{figure}[t]
  \centering
  \includegraphics[width=0.9\columnwidth]{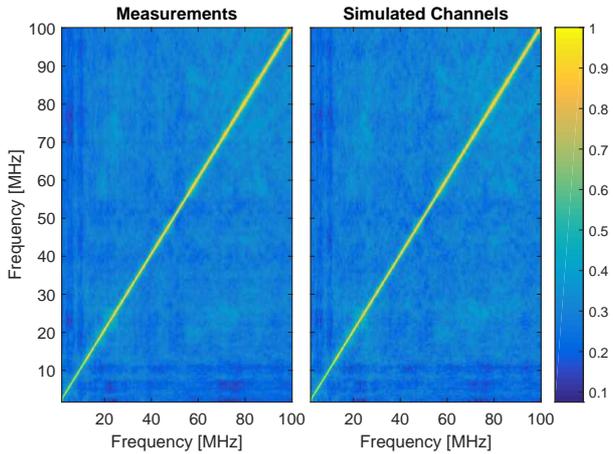}
  \caption{Normalized covariance matrix comparison for the experimental and the simulated SISO CFRs in dB scale.}
  \label{fig:CorrMtx}
\end{figure}
From the experimental measurements analysis it has been demonstrated that the amplitude $A(f)$ is log-normally distributed \cite{USAdb}, while the phase is uniformly distributed \cite{ESPdb,TCOMp1}. However, these two quantities exhibit a correlation between the different frequencies for each considered measurement. In order to simplify the problem the logarithm version of the CFR is considered, which is defined as $H_{dB}(f)=\log[H(f)]=A_{dB}(f)+j\varphi(f)$. It is straightforward to note as $A_{dB}$ is normally distributed, while $\varphi(f)$ is still uniformly distributed.

Thus, in order to account for the correlation in frequency, the normalized covariance matrix of the CFR, normalized by the product of the standard deviations of the considered variables, is considered. After some algebra, it turns out that the normalized covariance matrix (named $R_{H_{dB}}$) is given by $R_{H_{dB}}=R_{A_{dB}}+R_{\varphi}+i[R_{A_{dB},\varphi}^T-R_{A_{dB},\varphi}]$, where $\{\cdot\}^T$ is the transpose operator, while $R_{A_{dB}}$, $R_{\varphi}$ and $R_{A_{dB},\varphi}$ are the auto-covariance of the amplitude in dB, the phase and the covariance among the amplitude in dB and the phase, respectively. From the measurements analysis, it has been noted as the imaginary part of $R_{H_{dB}}$ is approximately zero (having an average value of 0.024), as the average value of $R_{A_{dB},\varphi}$ (nearly 0.033 on average). Hence, $A_{dB}$ and $\varphi(f)$ can be assumed uncorrelated (hypothesized as independent for simplicity), while $R_{H_{dB}}$ can be assumed as real and equal to the sum of $R_{A_{dB}}$ and $R_{\varphi}$.

The basic idea is to generate and correlate the amplitude (in dB scale) and the phase, independently, in order to obtain a set of CFRs that is statistically equivalent to the experimental measurements and with the same normalized covariance matrix, as clarified in Fig.~\ref{fig:CorrMtx}. The generation of a vector of correlated normal variables is simple and is given by $\mathbf{A}_{dB}^{cor}=(K_{A_{dB}})^{1/2}\mathbf{A}_{dB}^{ind}+\mathbf{m}$, where $\mathbf{A}_{dB}^{ind}\sim\mathcal{N}(0,\mathbf{I})$ is the vector of independent normal variables having mean $\mathbf{m}$, while $K_{A_{dB}}$ is the covariance matrix of the measurements amplitude. For the phase the process is a bit more complicated. In order to correlate uniform random variables (r.v.), the relation between the Pearson (linear) and the Spearman (rank) correlation is exploited \cite{Tonello_12, Versolatto_11, Tonello_10}. Given the target normalized covariance matrix $R_{\varphi}$ and the normal distribution function $N(\cdot)$, firstly a vector $\mathbf{x}$ of normal r.v. with a modified normalized covariance matrix $\hat{R}_{\varphi}=2\sin(R_{\varphi}\pi/6)$ is generated. Secondly, the vector of correlated uniform r.v. is obtained as $\mathbf{u}=N(\mathbf{x})$. As it can be noted in Fig.~\ref{fig:CorrMtx}, there is a good matching among the normalized covariance matrix of the measurements and that of the simulated channels. The only slight deviations are limited to the sharp transition areas, i.e. from high to low correlation values, or vice versa.

\subsection{Extension to the MIMO Scenario}
The exploitation of all the available wires, through a MIMO transmission technique in an extended frequency range \cite{HPAV2}, is of primarily interest. This, together with precoding, adaptation and a number of advanced techniques at the MAC layer, leads to a great performance improvement, enabling the achievement of much higher data transmission rates, in the order of 2 Gbps of peak rate. Thus, it is of interest to extend the SISO synthetic channel model discussed in Section~\ref{SISOmodel} to the MIMO scenario.

Similarly to the SISO case, the overall MIMO database of measurements can be represented through a singular four-dimensional matrix $\mathbf{H}$, having dimensions $N_M \times N_R \times N_T \times M$. In particular, $N_M$, $N_R$, $N_T$ and $M$ represent the number of measurements, receiving modes, transmitting modes and frequency samples, respectively. This matrix can be properly reshaped in order to consider not only the frequency correlation, but also the spatial correlation, existing among all the spatial mode combinations. In particular, the reshaped matrix $\mathbf{H}_r$ has dimensions $N_M \times N_R \cdot N_T \cdot M$. Each column of $\mathbf{H}_r$ represents all the measurements corresponding to a given frequency and referred to a specific transmitting-receiving mode combination. Hence, according to the same procedure discussed in Section~\ref{SISOmodel}, the covariance matrix among each column of $\mathbf{H}_r$ can be computed and then used in order to obtain an equivalent set of numerically simulated MIMO channels with a given target covariance. This strategy has been adopted for the amplitude only.

Concerning the phase, a different generation procedure is adopted. Indeed, the implementation method detailed in Section~\ref{SISOmodel} can be used for the emulation of the overall phase correlation exhibited by the entire database of measurements, especially among the different transmitting-receiving modes. However, in the following we consider a simpler approach. In particular, the unwrapped phase slope, along frequency, of the MIMO measurements has been assessed and statistically characterized. Then, an equivalent set of unwrapped phase profiles is computed imposing a slope that is randomly generated relying on the experimental slope distribution. As it will be discussed in the following section, this procedure allows to achieve good results, especially in terms of channel dispersion, despite its simplicity.

\subsection{Numerical Results}
\begin{figure}[t]
  \centering
  \includegraphics[width=0.9\columnwidth]{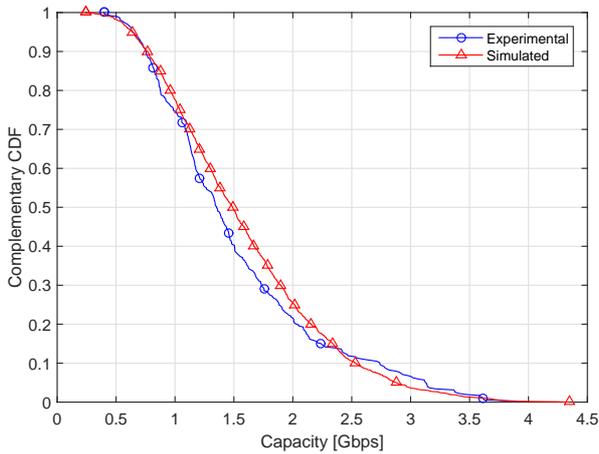}
  \caption{Capacity C-CDF comparison between experimental and simulated MIMO channels in the 1.8--100 MHz frequency band.}
  \label{fig:CCDFcmp}
\end{figure}
In order to validate the proposed synthetic MIMO channel model as an effective tool, which is able to emulate a typical PLC network, the distribution of the achievable performance is assessed. The MIMO channel measurements performed by the European Telecommunications Standards Institute (ETSI) special task force (STF) 410, discussed in \cite{ETSI_3}, are considered.
The focus is on the $2\times3$ MIMO channel, thus limiting the analysis on the three linearly independent star-style receiving modes, namely, phase (P), neutral (N) and common mode (CM), while feeding two $\Delta$-style transmitting modes. Concerning the noise, the MIMO PSD measurements provided by the STF-410 in \cite{ETSI_3} are assumed, considering also the noise correlation among the receiving modes, as discussed in \cite{ISPLC_14}.
In Fig.~\ref{fig:CCDFcmp} the capacity complementary cumulative distribution function (C-CDF), achieved by the 353 MIMO measurements and by an equivalent set of 2000 simulated channel realizations, is depicted. It can be noted as, apart from some small differences, mainly focused in the middle part, the distribution of the performance is practically the same for both the considered data sets. Indeed, the maximum discrepancy is in the order of 0.18 Gbps. The differences are mainly due to some inaccuracies in the generation process, as well as to some approximations, such as the independence of amplitude and phase discussed in Section~\ref{SISOmodel}.

Furthermore, also the average value of the most commonly used statistical metrics, namely the average channel gain (ACG), the root-mean-square delay spread (RMS-DS), the coherence bandwidth (CB) and the channel capacity, has been computed for both the experimental and simulated channels. The average is performed over the ensemble of realizations and among the different spatial modes. The computed quantities are listed in Table~\ref{tab:AVGvalMIMOmodel}. It can be noted as there is an almost perfect match among the metrics computed relying on the measurements and those calculated using the simulated channels, for all the considered quantities. The difference in terms of average capacity is approximately 3 \%, hence, a very low value. This results show as the proposed modeling strategy is able to capture all the properties and the relationships exhibited by a real communication scenario, despite its simplicity. In particular, it has been proved that it is more convenient to take into account a punctual, i.e. realization by realization, phase dependency along frequency, rather than impose a faithful overall spatial and frequency correlation for all the ensemble of realizations.
\begin{table}
\centering
\caption{\label{tab:AVGvalMIMOmodel}Average value of the main statistical metrics for both the experimental and the simulated MIMO channels.}
\begin{tabular}{c|c|c|c|c|c}
  \multirow{2}*{Type} & Band & $\overline{\mathcal{G}}$ & $\overline{\sigma}_\tau$ & $\overline{\mathcal{B}}_C^{(0.9)}$ & $\overline{\mathcal{C}}$ \\
                      & (MHz)& (dB) & ($\mu$s) & (kHz) & (Gbps) \\
  \hline
  Experimental & 1.8--100 & $-42.30$ & $0.350$ & $293.22$ & $1.53$ \\
  Simulated    & 1.8--100 & $-40.46$ & $0.353$ & $214.50$ & $1.64$ \\
\end{tabular}
\end{table}

A representative database of 100 MIMO PLC channel responses generated according to the proposed model described is made available online in \cite{ChDbase}.

\section{Conclusions}
This paper proposes an extremely synthetic top-down channel model for the MIMO PLC channel. The basic model idea and its theoretical basis have been discussed for the SISO case first. Then, the modeling procedure has been extended and tested on a specific in-home MIMO PLC scenario, relying on experimental channels. The proposed model has been evaluated comparing the main average statistical metrics and the performance, assessed in terms of capacity distribution. The strengths of the proposed model are the simplicity, relying only on the amplitude and phase properties, the flexibility, being suitable for different application contexts, and the statistical representativeness, as the results show. Indeed, the provided general model has the ability to emulate several PLC scenarios only by considering their specific properties. Thus, an interesting topic is the application of the proposed approach to other PLC application domains.

%\appendices
%\section{}
%Appendixes, if needed, appear before the acknowledgment.

%\section*{Acknowledgment}
%The authors would like to thank. Sponsor and financial support acknowledgment goes here.

%\bibliographystyle{IEEEtran}
%\bibliography{IEEEabrv,mybibliography}

\begin{thebibliography}{10}
\providecommand{\url}[1]{#1}
\csname url@samestyle\endcsname
\providecommand{\newblock}{\relax}
\providecommand{\bibinfo}[2]{#2}
\providecommand{\BIBentrySTDinterwordspacing}{\spaceskip=0pt\relax}
\providecommand{\BIBentryALTinterwordstretchfactor}{4}
\providecommand{\BIBentryALTinterwordspacing}{\spaceskip=\fontdimen2\font plus
\BIBentryALTinterwordstretchfactor\fontdimen3\font minus
  \fontdimen4\font\relax}
\providecommand{\BIBforeignlanguage}[2]{{%
\expandafter\ifx\csname l@#1\endcsname\relax
\typeout{** WARNING: IEEEtran.bst: No hyphenation pattern has been}%
\typeout{** loaded for the language `#1'. Using the pattern for}%
\typeout{** the default language instead.}%
\else
\language=\csname l@#1\endcsname
\fi
#2}}
\providecommand{\BIBdecl}{\relax}
\BIBdecl

\bibitem{HPAV2}
L.~Yonge, J.~Abad, K.~Afkhamie, L.~Guerrieri, S.~Katar, H.~Lioe, P.~Pagani,
  R.~Riva, D.~M. Schneider, and A.~Schwager, ``{An Overview of the HomePlug AV2
  Technology},'' \emph{J. Elect. Comput. Eng.}, vol. 2013, pp. 1--20, 2013.

\bibitem{IEEE_P1901}
``{IEEE Standard for Broadband over Power Line Networks: Medium Access Control
  and Physical Layer Specifications},'' {IEEE 1901-2010}, September 2010.

\bibitem{ITU_Ghn}
\BIBentryALTinterwordspacing
``{Unified High-Speed Wireline-Based Home Networking Transceivers -- System
  Architecture and Physical Layer Specification},'' {Recommendation ITU-T
  G.9960}, {ITU}, December 2010. [Online]. Available:
  \url{https://www.itu.int/rec/T-REC-G.9960/en}
\BIBentrySTDinterwordspacing

\bibitem{BottomUp}
T.~Esmailian, F.~R. Kschischang, and P.~{Glenn Gulak}, ``{In-Building Power
  Lines as High-Speed Communication Channels: Channel Characterization and a
  Test Channel Ensemble},'' \emph{Intern. J. of Commun. Syst.}, vol.~16, no.~5,
  pp. 381--400, June 2003.

\bibitem{BU_Ton}
A.~M. Tonello and F.~Versolatto, ``{Bottom-Up Statistical PLC Channel Modeling
  -- Part I: Random Topology Model and Efficient Transfer Function
  Computation},'' \emph{IEEE Trans. Power Del.}, vol.~26, no.~2, pp. 891--898,
  April 2011.

\bibitem{MPM}
M.~Zimmermann and K.~Dostert, ``{A Multipath Model for the Powerline
  Channel},'' \emph{IEEE Trans. Commun.}, vol.~50, no.~4, pp. 553--559, April
  2002.

\bibitem{TD_Ton}
A.~M. Tonello, ``{Wideband Impulse Modulation and Receiver Algorithms for
  Multiuser Power Line Communications},'' \emph{EURASIP J. on Advances in
  Signal Processing}, vol. 2007, pp. 1--14, 2007.

\bibitem{IEICE}
A.~M. Tonello, A.~Pittolo, and M.~Girotto, ``{Power Line Communications:
  Understanding the Channel for Physical Layer Evolution Based on Filter Bank
  Modulation},'' \emph{IEICE Trans. Commun.}, vol. E97-B, no.~8, pp.
  1494--1503, August 2014.

\bibitem{USAdb}
S.~Galli, ``{A Novel Approach to the Statistical Modeling of Wireline
  Channels},'' \emph{IEEE Trans. Commun.}, vol.~59, no.~5, pp. 1332--1345, May
  2011.

\bibitem{ESPdb}
J.~A. Cort{\'e}s, F.~J. Ca{\~n}ete, L.~D{\'i}ez, and J.~L.~G. Moreno, ``{On the
  Statistical Properties of Indoor Power Line Channels: Measurements and
  Models},'' in \emph{Proc. Int. Symp. Power Line Commun. and Its App.
  (ISPLC)}, Udine, Italy, April 2011, pp. 271--276.

\bibitem{TCOMp1}
A.~M. Tonello, F.~Versolatto, and A.~Pittolo, ``{In-Home Power Line
  Communication Channel: Statistical Characterization},'' \emph{IEEE Trans.
  Commun.}, vol.~62, no.~6, pp. 2096--2106, June 2014.

\bibitem{Tonello_12}
A.~M. Tonello, F.~Versolatto, B.~Bejar, and S.~Zazo, ``{A Fitting Algorithm for
  Random Modeling the PLC Channel},'' \emph{IEEE Transactions on Power
  Delivery}, vol.~27, no.~3, pp. 1477--1484, July 2012.

\bibitem{Versolatto_11}
F.~Versolatto and A.~M. Tonello, ``{An MTL Theory Approach for the Simulation
  of MIMO Power-Line Communication Channels},'' \emph{IEEE Transactions on
  Power Delivery}, vol.~26, no.~3, pp. 1710--1717, July 2011.

\bibitem{Tonello_10}
A.~M. Tonello and F.~Versolatto, ``{Bottom-Up Statistical PLC Channel Modeling
  -- Part II: Inferring the Statistics},'' \emph{IEEE Transactions on Power
  Delivery}, vol.~25, no.~4, pp. 2356--2363, Oct 2010.

\bibitem{ETSI_3}
{ETSI TR 101 562-3 V 1.1.1}, ``{PowerLine Telecommunications (PLT); MIMO PLT;
  Part 3: Setup and Statistical Results of MIMO PLT Channel and Noise
  Measurements},'' European Telecommunication Standardization Institute, Tech.
  Rep., 2012.

\bibitem{ISPLC_14}
A.~Pittolo, A.~M. Tonello, and F.~Versolatto, ``{Performance of MIMO PLC in
  measured channels affected by correlated noise},'' in \emph{Proc. of IEEE
  Int. Symp. on Power Line Commu. and its App. (ISPLC)}, March 2014, pp.
  261--265.

\bibitem{ChDbase}
\BIBentryALTinterwordspacing
``{Synthetic MIMO PLC Channel Database},'' April 2016. [Online]. Available:
  \url{http://www.andreatonello.com}
\BIBentrySTDinterwordspacing

\end{thebibliography}

\end{document}